\documentclass[floatfix,twocolumn,superscriptaddress,longbibliography,bibnotes]{revtex4-1}
\usepackage{physics}
\usepackage{graphicx}
\usepackage{xcolor}
\usepackage[pdftex,bookmarks=true,bookmarksopen,bookmarksnumbered,
                colorlinks,
                linkcolor=blue,
                citecolor=blue]{hyperref}
                
\usepackage{cleveref}
\usepackage{siunitx}
\usepackage{amsmath}
\usepackage[version=4]{mhchem}
\usepackage{colortbl}
\usepackage{float}
\usepackage{siunitx}
\usepackage{newfloat}
\usepackage{pgfplots}
\usepackage{upgreek}
\pgfplotsset{compat=newest}
\usepackage{standalone}
\usepackage{upgreek}

\begin{document}

\title{Spin thermometry and spin relaxation of optically detected \texorpdfstring{Cr$\mathbf{^{3+}}$}{Cr3+} ions in \texorpdfstring{Al$\mathbf{_2}$O$\mathbf{_3}$}{Al2O3} (ruby)}

\author{Vikas K. Sewani}
\email[]{v.sewani@student.unsw.edu.au}
\affiliation{
Centre for Quantum Computation and Communication Technology, School of Electrical Engineering and Telecommunications, UNSW Sydney, Sydney, New South Wales 2052, Australia
}

\author{Rainer J. St\"{o}hr}
\affiliation{
 $3^{rd}$ Physikalisches Institut, Universit\"{a}t Stuttgart, 70569 Stuttgart, Germany
}

\author{Roman Kolesov}
\affiliation{
 $3^{rd}$ Physikalisches Institut, Universit\"{a}t Stuttgart, 70569 Stuttgart, Germany
}

\author{\\Hyma H. Vallabhapurapu}
\affiliation{
Centre for Quantum Computation and Communication Technology, School of Electrical Engineering and Telecommunications, UNSW Sydney, Sydney, New South Wales 2052, Australia
}

\author{Tobias Simmet}
\affiliation{
 Walter Schottky Institut and Physik Department, Technische Universit\"{a}t M\"{u}nchen, 85748 Garching, Germany
}

\author{Andrea Morello}
\affiliation{
Centre for Quantum Computation and Communication Technology, School of Electrical Engineering and Telecommunications, UNSW Sydney, Sydney, New South Wales 2052, Australia
}

\author{Arne Laucht}
\email[]{a.laucht@unsw.edu.au}
\affiliation{
Centre for Quantum Computation and Communication Technology, School of Electrical Engineering and Telecommunications, UNSW Sydney, Sydney, New South Wales 2052, Australia
}

\begin{abstract}
Paramagnetic ions in solid state crystals form the basis for many advanced technologies such as lasers, masers, frequency standards, and quantum-enhanced sensors. One of the most-studied examples is the $\rm{Cr}^{3+}$ ion in sapphire ($\rm{Al}_2\rm{O}_3$), also known as ruby, which has been intensely studied in the 1950s and 1960s. However, despite decades of research on ruby, some of its fundamental optical and spin properties have not yet been characterized at ultra low-temperatures.
In this paper, we present optical measurements on a ruby crystal in a dilution refrigerator at ultra-low temperatures down to 20 mK. Analyzing the relative populations of its $\rm{^4A_2}$ ground state spin levels, we extract a lattice temperature of $143\pm7$~mK under continuous laser excitation. We perform spin lattice relaxation $T_1$ measurements in excellent agreement with the direct, one-phonon model. Furthermore, we perform optically detected magnetic resonance measurements showing magnetically driven transitions between the ground state spin levels for various magnetic fields.
Our measurements characterize some of ruby's low temperature spin properties, and lay the foundations for more advanced spin control experiments. 
\end{abstract}

\maketitle

\section{Introduction}

In addition to being a popular gemstone, ruby ($\mathrm{Cr}^{3+}$ in $\mathrm{Al}_{2}\mathrm{O}_{3}$) has also been investigated for its optical and spin properties for over 150 years~\cite{Becquerel1867,Imbusch1987}, and has played an important role in the history of lasers~\cite{Maiman1960} and masers~\cite{Makhov1958}. However, despite decades of research, it has rarely been measured at temperatures $T<1$~K~\cite{Farr2013,Farr2014,Wiemann2015,Miksch2020}, and fundamental properties, such as its electron spin relaxation, need to be investigated. In particular, the spin relaxation of the $3\mathrm{d}^{3}$ orbital ground state of $\mathrm{Cr}^{3+}$, which consists of an $\mathrm{S}=3/2$ Kramer's doublet with a zero-field splitting of $\sim11.49$~GHz (corresponding to the thermal energy at $T=550$~mK), has only been measured at temperatures as low as $T=1.6$~K, where a maximum relaxation time of $T_{1} = 560$~ms was measured~\cite{Standley1965}. For ruby crystals with sufficiently low $\mathrm{Cr}^{3+}$ concentrations ($<0.01$\%), the spin relaxation follows a $T_{1}(T)\propto T^{-1}$ relationship down to this temperature, corresponding to a one-phonon (direct) process~\cite{Standley1965,Donoho}. Assuming the one-phonon model continues to dominate at lower temperatures, a saturation of the  $T_{1}$ can only be expected below $T=550$ mK at zero external magnetic field, which has not previously been measured.

In this paper, we perform optical measurements on a ruby crystal inside a dilution refrigerator with base temperature $T_{\mathrm{MXC}}=20$~mK. By mapping out the population ratio of the Zeeman-split $|{\pm3/2}\rangle_{\mathrm{g}}$ spin states as a function of magnetic field, we conduct optical thermometry of the ruby lattice temperature, and measure $T_{\mathrm{Ruby}}=143\pm7$~mK under $15$~nW resonant laser excitation. This proves that a very low temperature of the ruby sample can be maintained under sufficiently low continuous laser excitation. We then measure the spin $T_{1}$ decay of the $\mathrm{Cr}^{3+}$ ground state using an all-optical method at ultra-low temperatures. We obtain a maximum $T_1=3.67\pm0.35$~s, and observe that the spin relaxation can be completely described by a direct, one-phonon process. 

Furthermore, we demonstrate optically-detected magnetic resonance (ODMR) of the $\mathrm{Cr}^{3+}$ ground state. Previously, electron paramagnetic resonance (EPR) spectroscopy has been utilized to probe the Zeeman level splitting of the $\rm{^{4}A_{2}}$ ground state in an EPR spectrometer~\cite{Manenkov1955,Bois1959,Manenkov1960}, and more recently via coupling to a microwave whispering gallery mode~\cite{Hartnett2007,Farr2013,Farr2014}, or a coplanar waveguide~\cite{Wiemann2015,Miksch2020}. Optically-detected readout has been shown to provide access to both the spin population of the optically excited $\rm{\Bar{E}(^{2}E)}$ state~\cite{Geschwind1959,Geschwind1965}, as well as of the $\rm{^{4}A_{2}}$ ground state through broadband optical pumping via spin memory when the external magnetic field is aligned with the c-axis of the crystal~\cite{Geschwind1959,Imbusch1966}. 
In our experiment, we use resonant laser excitation for spin initialization and readout~\cite{Koehl2017}, which improves our spatial resolution to $\sim1$~$\upmu$m$^2$ and strongly reduces the number of ions that are probed compared to EPR spectroscopy. Additionally, it allows the magnetic field to be applied non-parallel to the c-axis of the crystal, providing access to clock transitions~\cite{Bois1959}, where the coherence times of the spins are expected to be much longer. A printed circuit board (PCB) antenna, mounted in the vicinity of the sample, allows us to magnetically drive the ground state electron spins. While not demonstrated in this paper, the combination of resonant laser readout with a PCB or on-chip microwave (MW) antenna not only allows spectroscopic, but also coherence time investigations. This adds to the spin characterization tools available for the future study of the $\mathrm{Cr}^{3+}$ defect in ruby.

\section{Spin Hamiltonian}\label{sec_hamiltonian}

\begin{figure*} 
	\centering 
		\includegraphics[width=\textwidth]{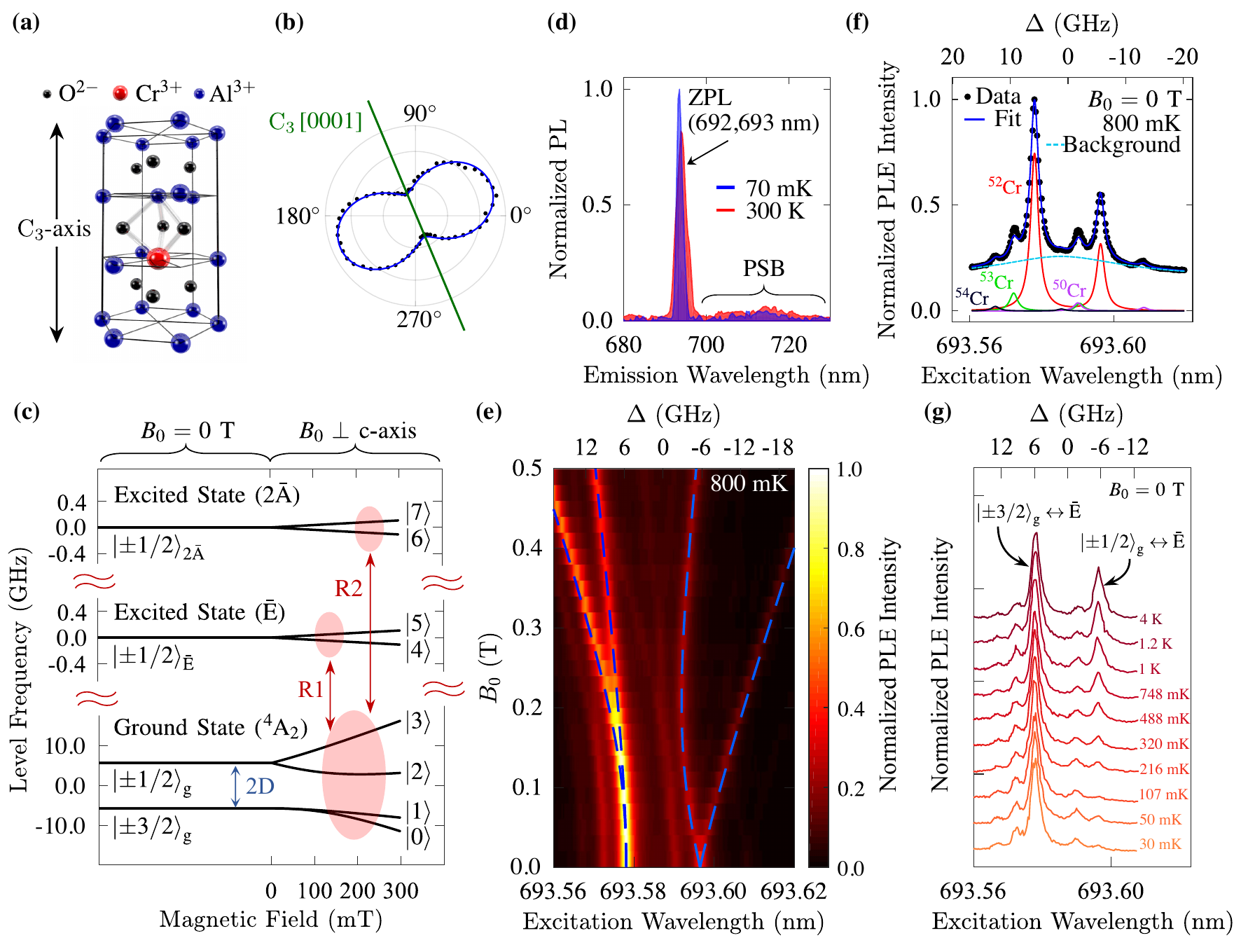}
		\caption{
		(a) Hexagonal crystal structure of $\mathrm{Al}_{2}\mathrm{O}_{3}$ containing a single interstitial $\mathrm{Cr}^{3+}$ ion. 
		(b) Polarization dependence of the phonon sideband (PSB) emission with excitation laser in resonance with the $\ket{\pm3/2}_{\mathrm{g}}\leftrightarrow\bar{\mathrm{E}}$ transition at $B_0=0$~T. 
		(c) Electronic level scheme of $\mathrm{Cr}^{3+}$ in $\mathrm{Al}_{2}\mathrm{O}_{3}$ at $B_{0}=0$ T and $B_{0}\perp$ c-axis.  
		(d) Emission spectrum of ruby with 520~nm  (off-resonant) excitation at room temperature (red curve) and $T_{\rm{MXC}}=70$~mK (blue curve). Marked are the zero-phonon lines (ZPL) used for resonant excitation and the PSB used for signal detection in all further experiments. 
		(e) PLE spectrum as a function of magnetic field for $B_{0}\perp$ c-axis. The overlaid, dashed, blue lines correspond to the Hamiltonian in Eq.~\ref{eqn:hamiltonian} with parameters $g=1.982$, $-2D=11.460$ GHz, and 693.587 nm as the splitting between the midpoint of $\ket{\pm1/2}_{\mathrm{g}}$ and $\ket{\pm3/2}_{\mathrm{g}}$, and $\bar{\mathrm{E}}$. The quantity $\Delta$ (top x-axis) denotes the detuning in laser excitation energy (in frequency units) from the mid-point of the $^{52}$Cr $\ket{\pm1/2}_{\mathrm{g}}\leftrightarrow\bar{\mathrm{E}}$ and $\ket{\pm3/2}_{\mathrm{g}}\leftrightarrow\bar{\mathrm{E}}$ transitions. 
		(f) Zero-field photoluminescence excitation (PLE) spectrum showing multiple peaks that are associated with the different $\mathrm{Cr}$ isotopes. Peaks are fitted with Lorentzian functions with their relative peak intensities fixed to correspond to the isotopes' natural abundance (4.3\% for \ce{^{50}Cr}, 83.8\% for \ce{^{52}Cr}, 9.5\% for \ce{^{53}Cr}, and 2.4\% for \ce{^{54}Cr}~\cite{Imbusch1964}).
		(g) PLE spectrum at $B_{0}=0$~T for varying mixing chamber (MXC) temperatures, where the temperature readings are given by the MXC thermometer. Increased PLE intensity from the $\ket{\pm1/2}_{\mathrm{g}}\leftrightarrow\bar{\mathrm{E}}$ transition is in accordance with the redistribution of spin population at increased temperature.}
		\label{figure_ple}
\end{figure*}

In ruby, some of the $\mathrm{Al}^{3+}$ ions (usually much less than 1\%) of the $\mathrm{Al}_{2}\mathrm{O}_{3}$ crystal are replaced by $\mathrm{Cr}^{3+}$ ions with three unpaired electrons in their outer 3$\mathrm{d}^{3}$ shell. Each $\mathrm{Cr}^{3+}$ ion is surrounded by six $\mathrm{O}^{2-}$ ions in the form of a distorted octahedron [see Fig.~\ref{figure_ple}(a)]~\cite{Imbusch1987}. The electrostatic crystal field, which the chromium ions are subject to, lift the 3d orbital degeneracy of the $^{4}\mathrm{A}_{2}$, $\bar{\mathrm{E}}$ and $2\bar{\mathrm{A}}$ levels, as shown in the energy level diagram in Fig.~\ref{figure_ple}(c). In the ground state ($^{4}\mathrm{A}_{2}$), the chromium ions are additionally subject to a trigonal crystal field, and together with spin-orbit coupling, the spin degeneracy is lifted, leaving a Kramer's doublet with the degenerate pairs $\ket{\pm1/2}$ and $\ket{\pm3/2}$, with the latter being the spin ground state~\cite{Powell1966,Bates1969}. At cryogenic temperatures, the higher excited orbital states $\bar{\mathrm{E}}$ and $2\bar{\mathrm{A}}$ are $1.7876$~eV ($693.6$~nm) and $1.792$~eV ($692.2$~nm) above $^{4}\mathrm{A}_{2}$. Due to their emission in the red part of the spectrum, they are referred to as the R1 and R2 lines, respectively~\cite{Imbusch1987}. Eq.~\ref{eqn:hamiltonian} shows the effective system Hamiltonian of the $^{4}\mathrm{A}_{2}$ ground state (${\cal H}_{0}$), and Eq.~\ref{eqn:hamiltonian_exc} shows the Hamiltonian of the $\bar{\mathrm{E}}$ (${\cal H}_{\bar{\mathrm{E}}}$) and $2\bar{\mathrm{A}}$ (${\cal H}_{2\bar{\mathrm{A}}}$) excited states, which are the same up to their $g$-factors:

\begin{equation}
    \label{eqn:hamiltonian}
    \begin{gathered}
    {\cal H}_0=g_{\parallel}{\mu_{\rm B}}{B}_{\rm  z}\mathrm{S}_{\rm z}+g_{\perp}{\mu_{\rm B}}({B}_{\rm x}\mathrm{S_{x}}+{B}_{\rm y}\mathrm{S_{y}}) +\\D[\mathrm{S_{z}^2-\frac{1}{3}S(S+1)}],
    \end{gathered}
\end{equation}

\begin{equation}
    \label{eqn:hamiltonian_exc}
    \begin{gathered}
    {\cal H}_{\bar{\mathrm{E}},2\bar{\mathrm{A}}}=g_{\parallel}{\mu_{\rm B}}{B}_{\rm  z}\mathrm{S}_{\rm z}+g_{\perp}{\mu_{\rm B}}({B}_{\rm x}\mathrm{S_{x}}+{B}_{\rm y}\mathrm{S_{y}}).
    \end{gathered}
\end{equation}

The z-axis for ${\cal H}_{0}$ and ${\cal H}_{\bar{\mathrm{E}},2\bar{\mathrm{A}}}$ is defined along the $\mathrm{Al}_{2}\mathrm{O}_{3}$ crystal axis (c-axis) direction, while the x-axis and y-axis are arbitrarily oriented. For ${\cal H}_{0}$ ($^{4}\mathrm{A}_{2}$), S=3/2 where ${\rm S}_{\rm  x,y,z}$ are the corresponding spin matrices, ${B}_{\rm  x,y,z}$ are the vector components of an externally applied magnetic field, ${\mu_{\rm B}}$ is the Bohr's magneton, $-2D=11.493\pm0.004$~GHz is the zero-field splitting, and $g_{\parallel}=1.9817\pm0.0004$ and $g_{\perp}=1.9819\pm0.0006$ are the room temperature $g$-factors parallel and perpendicular to the c-axis~\cite{Chang1978}. The Hamiltonians ${\cal H}_{\bar{\mathrm{E}}}$ and ${\cal H}_{2\bar{\mathrm{A}}}$ for the excited states consist of only Zeeman terms, where ${\rm S}_{\rm  x,y,z}$ are the S=1/2 spin matrices, $g_{\parallel}=-2.445$ and $g_{\perp}=0.0515$ for $\bar{\mathrm{E}}$~\cite{Muramoto1969}, and $g_{\parallel}=1.46$ and $g_{\perp}\approx0$ for $2\bar{\mathrm{A}}$~\cite{Sugano1958b,Hori1979}. 

In our experiments, we fix $B_{0}\perp\rm{C}_3 [0001]$-axis. Additionally, our optics axis is parallel to $B_{0}$, which allows us to determine the orientation of the $\rm{C}_3 [0001]$ axis with respect to our optical setup with a laser polarization dependence measurement, as shown in Fig.~\ref{figure_ple}(b)~\cite{Nelson1965,Powell1966}. The energy levels as a function of magnetic field for $B_{0}\perp$ c-axis are shown in Fig.~\ref{figure_ple}(c). For finite values of $B_{0}$, ${\cal H}_{0}$ contains off-diagonal terms and hence does not commute with the $\mathrm{S_{z}}$ spin matrix. This results in the eigenstates consisting of superpositions of the pure spin S=3/2 eigenstates with their quantization along the $\mathrm{C}_{3}$ symmetry axis. We hence label all states canonically from $\ket{0}-\ket{7}$ as shown. See also Ref.~\cite{Bois1959} for a detailed study of the magnetic field direction dependence of the eigenvalues and eigenstates of ${\cal H}_{0}$.

\section{Experimental Setup}\label{sec_setup}

We perform our measurements on a mono-crystalline ruby sample with a $\mathrm{Cr}^{3+}$ concentration of 0.005$\%$. For optical excitation, we either use a 520~nm laser-diode for off-resonant excitation, or a continuous wave titanium-sapphire (Ti:Sa) laser, tunable between 692 - 1000~nm with optical linewidth $<50$~kHz and referenced with a wavemeter, for resonant excitation of the R1 line. In either case, the collimated beam enters the dilution refrigerator from the top through a vacuum-sealed, anti-reflective window, and propagates down the cryostat through small apertures ($D=7$~mm) in the baffles of the 50K, 4K and 1K plates. The laser beam is incident on an objective lens (Olympus MS Plan 50x/0.80NA), which is mounted to the cold finger attached to the mixing chamber (MXC) plate. The laser beam then gets focused through a 1~mm hole in a printed circuit board (PCB) that carries the MW antenna~\cite{Sewani2020}, which delivers the oscillating magnetic field $B_{1}{\parallel}B_{0}$ for the ODMR experiments in Sec.~\ref{sec_odmr}. The laser is focused onto the ruby sample, which is mounted on top of a 3-axis \textit{Attocube} stage on the same cold finger, and thermalized through an \textit{Attocube} thermal coupling link. The ruby sample is positioned in the homogeneous region at the center of a superconducting electromagnet, which delivers the static magnetic field $B_{0}$. The sample's c-axis is perpendicular to the $B_{0}$ direction ($B_{0}\perp$ c-axis).

The photoluminescence (PL) from an ensemble of $\mathrm{Cr}^{3+}$ ions is collimated by the same objective lens, and travels through the same path as discussed before, for detection outside the fridge. The PL signal is coupled into a 50~$\upmu$m multi-mode fiber, and detection is performed with either a mini-spectrometer or a single-photon avalanche diode (SPAD). The PL spectrum obtained with off-resonant excitation is shown in Fig.~\ref{figure_ple}(d) at room temperature (red line) and $T_{\rm{MXC}}=70$~mK (blue curve). The zero phonon line (ZPL) is clearly visible in both cases, resulting from PL from both the R1 and R2 lines. Additionally, a wide phonon sideband (PSB) is visible on the low-energy side of the ZPL. For all further measurements, we resonantly excite the R1 line, and detect the PL from the PSB by placing a 700~nm long-pass and a 800~nm short-pass filter into the detection path.

In order to generate the control sequences and synchronize the different pieces of equipment, we use a transistor-transistor logic (TTL) pulse generator (\textit{PulseBlasterESR-PRO} 250). The output of the Ti:Sa laser is amplitude modulated by the TTL pulses through an acousto-optic modulator. The MW drive, delivered by the PCB antenna, is modulated via the built-in pulse modulation of the MW source. Finally, the digitizer used to record the single-photon detection events from the SPAD is gated, allowing us to temporally select when we would like the digitizer to store and count pulses with respect to the laser and MW pulses.

\section{Photoluminescence excitation spectroscopy}\label{sec_photo}
We start by employing photoluminescence excitation (PLE) spectroscopy to determine the R1 optical absorption lines as a function of $B_{0}$. When the Ti:Sa laser is in resonance with an absorption line, the excitation rate from the respective ground state level is increased, resulting in a increased PL from both the ZPL and PSB. At $B_{0}=0$ T, we only expect to see two optical transitions originating from $\ket{\pm1/2}_{\mathrm{g}}\leftrightarrow\bar{\mathrm{E}}$ and $\ket{\pm3/2}_{\mathrm{g}}\leftrightarrow\bar{\mathrm{E}}$, while at $B_{0}>0$, a total of 4 transitions can be observed within the R1 line [see also Fig.~\ref{figure_ple}(c)]. We plot the PLE spectra in Fig.~\ref{figure_ple}(e) up to $B_0=0.5$~T. The measurement is performed at a temperature ($T_{\rm{MXC}}=800$~mK)~$>$~($2D\hbar/\mathrm{k_{B}}=550$~mK) to allow for sufficient thermal population of the $\ket{\pm1/2}_{\mathrm{g}}$ state. Up to this magnetic field, we expect the Zeeman splitting of $\bar{\mathrm{E}}$ to be $<360$~MHz owing to the small value of $g_{\perp}=0.0515$ (see also Eq.~\ref{eqn:hamiltonian_exc}). As this is much smaller than the optical linewidths of the PLE transitions ($\sim$1.8 GHz), the Zeeman splitting of $\bar{\mathrm{E}}$ cannot be observed, and the ground state Hamiltonian ${\cal H}_{0}$ dominates the PLE spectra. The most prominent transitions in Fig.~\ref{figure_ple}(e) are in good agreement with Eq.~\ref{eqn:hamiltonian}, and can be attributed to the R1 transitions of the $\mathrm{^{52}Cr}$ isotope, based on its natural abundance of 83.8$\%$. Other isotopes of chromium result in a blue shift or red shift of the R1 lines, and hence appear as dimmer copies near the dominant lines, as seen in Figs.~\ref{figure_ple}(e,f). Of these isotopes, $^{53}\rm{Cr}$ carries a nuclear spin of $I=3/2$ which couples to the electron spin through a hyperfine interaction of $A=48.5$~MHz~\cite{Terhune1961}.

In order to probe the effect of temperature on the relative strengths of the absorption lines, we vary the temperature of the dilution refrigerator and plot the measured PLE spectra at $B_0=0$~T in Fig.~\ref{figure_ple}(g). The MXC temperature $T_{\rm{MXC}}$ is altered via a resistive heater mounted on the MXC plate, and is measured by the MXC thermometer. At lower temperatures, $T_{\rm{MXC}}$ is not a good measure of the lattice temperature near the $\rm{Cr}^{3+}$ ions under investigation. This is due to (i) the imperfect thermalization of the ruby sample to the MXC, (ii) local heating due to laser irradiation, and (iii) limited cooling power of the fridge. However, quenching of the $\mathrm{\pm\ket{1/2}_{g}}\leftrightarrow\bar{\mathrm{E}}$ transition and greater spin polarization in the $\pm\mathrm{\ket{3/2}_{g}}$ state for $T_{\rm{MXC}}<500$~mK in Fig.~\ref{figure_ple}(g) confirm that a lattice temperature $<550$~mK is achieved. We will determine the exact lattice temperature by conducting optical thermometry on the $\rm{Cr}^{3+}$ ions in the next section (Sec.~\ref{sec_thermometry}).

\begin{figure}[!t]
	\centering 
		\includegraphics[width=\columnwidth]{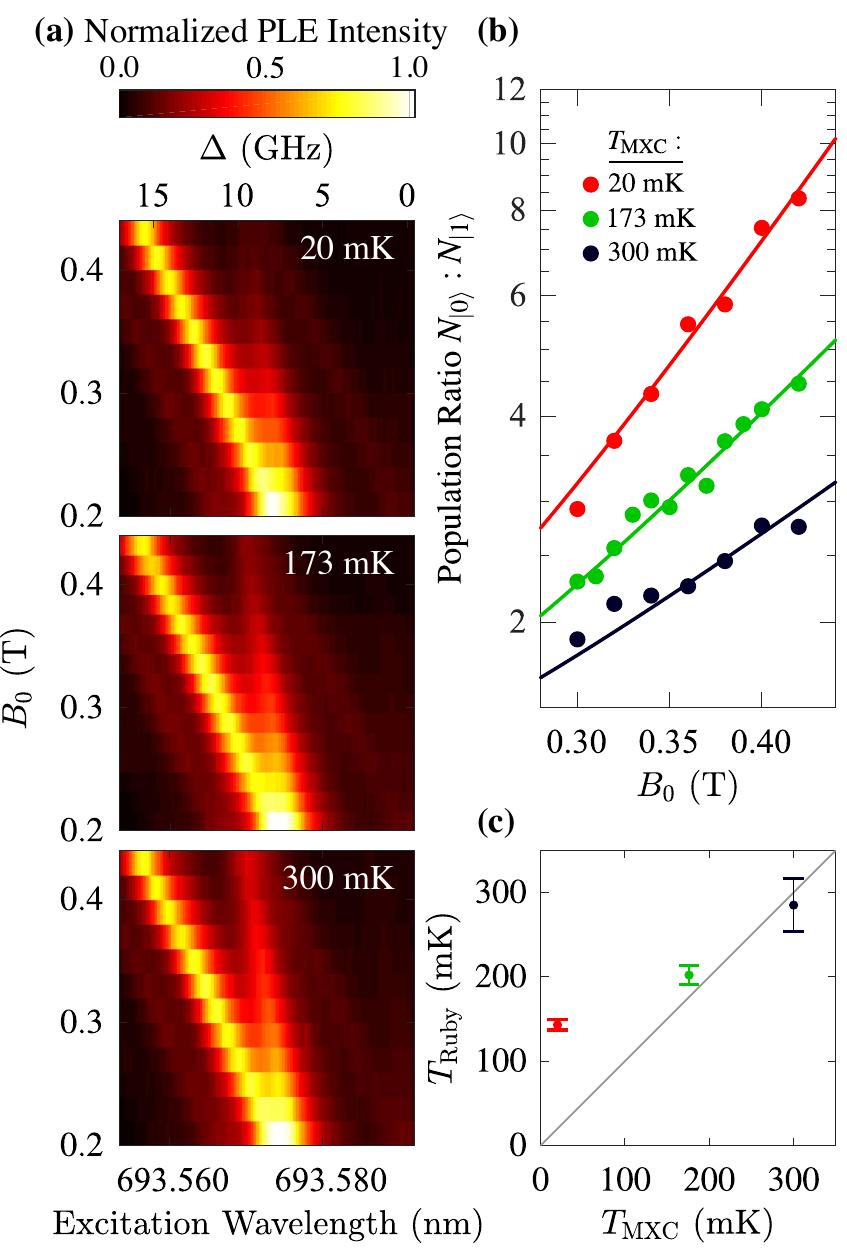}
		\caption{
		(a) Normalized PLE spectra of the $\ket{0}\leftrightarrow\bar{\mathrm{E}}$ and $\ket{1}\leftrightarrow\bar{\mathrm{E}}$ transitions at $T_{\mathrm{MXC}} = 20$, 173 and 300~mK. 
		(b) Ratio of PLE peak intensities (circles) extracted from (a) as a function of magnetic field. The solid lines are fits to the Boltzmann distribution, determining $T_{\mathrm{Ruby}}=143\pm7$, $202\pm12$, and  $285\pm31$~mK for $T_{\mathrm{MXC}}=20$, $173$, and $300$~mK, respectively. 
		(c) Extracted $T_{\rm{Ruby}}$ vs. $T_{\rm{MXC}}$.
		}
		\label{figure_thermometry}
\end{figure}

\section{Optical Thermometry}\label{sec_thermometry}
For the measurements shown in Figure~\ref{figure_ple}(g), we could only state the temperatures obtained from the MXC thermometer $T_{\rm{MXC}}$, which most certainly differ from the actual lattice temperatures $T_{\mathrm{Ruby}}$. While $T_{\mathrm{Ruby}}$ can be inferred from $\mathrm{Cr}^{3+}$ spin relaxation measurements for temperatures $T_{\mathrm{MXC}}>550$~mK (see Sec.~\ref{sec_t1}), we can also extract $T_{\mathrm{Ruby}}$ from the relative spin populations of any two level system in the ground state with splitting $\sim\mathrm{k_B}T_{\mathrm{Ruby}}$. For the measurements in this section (see Fig.~\ref{figure_thermometry}), we choose the $\ket{\pm3/2}_{\mathrm{g}}$ states, that split into $\ket{0}$ and $\ket{1}$ subject to a magnetic field [see also Sec.~\ref{sec_hamiltonian} and Fig.~\ref{figure_ple}(c)]. At $B_{0}=0.3$ T, the $\ket{0}$ and $\ket{1}$ states can be clearly resolved, and their level splitting is 3.46~GHz (from Eq.~\ref{eqn:hamiltonian}) which corresponds to the thermal energy at $T=166$~mK. We determine their relative populations by performing a PLE scan similar to Fig.~\ref{figure_ple}(e), where the PL intensity is proportional to the spin population. A laser excitation power of 15~nW is used to avoid the effects of optical cycling, which would deviate the electron populations away from their thermal distributions (see Appendix~\hyperref[App_C]{C}). This is in contrast to the $T_{1}$ measurements performed in Sec.~\ref{sec_t1}, where a 1~$\upmu$W laser excitation power is required to deviate the system sufficiently from thermal equilibrium, enabling us to measure the decay of the spin population over the relaxation duration. 

The results of the optical thermometry method described above  are shown in Fig.~\ref{figure_thermometry}(a), where we plot the PLE spectra of the $\ket{0}\leftrightarrow\bar{\mathrm{E}}$ and $\ket{1}\leftrightarrow\bar{\mathrm{E}}$ transitions as a function of $B_0$ for $T_{\mathrm{MXC}}=20$, 173, 300~mK. In Fig.~\ref{figure_thermometry}(b), we extract the PL intensity ratio of these two transitions -- which is proportional to the population ratio of $\ket{0}$ and $\ket{1}$ -- for $0.3 \leq B_{0} \leq 0.42$~T, and fit this ratio to the Boltzmann distribution,
\begin{equation}\label{eqn:boltzmann}
    \frac{N_{\ket{0}}(B_0)}{N_{\ket{1}}(B_0)} = \mathrm{exp}\left(\frac{E_{\ket{1}}(B_0)-E_{\ket{0}}(B_0)}{{\rm k}_{\rm B}T_{\rm{Ruby}}}\right),    
\end{equation}
where $E_{\ket{0},\ket{1}}(B_0)$ are the energies of the states $\ket{0}$, $\ket{1}$ as a function of $B_0$, obtained from Eq.~\ref{eqn:hamiltonian}. We find that at $T_{\mathrm{MXC}}=20$ mK the lattice temperature $T_{\mathrm{Ruby}}=143\pm7$~mK is significantly higher, due to local heating from the laser and insufficient thermalization of the sample. Nonetheless, this measurement verifies that lattice temperatures $<150$~mK are achievable in optical experiments under continuous laser excitation, similar to what has been demonstrated with self-assembled quantum dots in Ref.~\cite{Haupt2014}. At $T_{\mathrm{MXC}}=173$~mK, we measure $T_{\mathrm{Ruby}}=202\pm12$~mK, and at $T_{\mathrm{MXC}}=300$~mK, we measure $T_{\mathrm{Ruby}}=285\pm31$~mK [data summarized in Fig.~\ref{figure_thermometry}(c)]. These results show that, under these experimental conditions, the ruby sample is fully thermalized with the MXC at $T_{\mathrm{MXC}} \approx  300$~mK.

\section{Spin-lattice relaxation}\label{sec_t1}
We now measure the temperature dependence of the spin relaxation time $T_{1}$ from the $\ket{\pm1/2}_{\mathrm{g}}$ state to the $\ket{\pm3/2}_{\mathrm{g}}$ state at $B_{0}=0$~T. At thermal equilibrium, we expect the relative populations of the sublevels of the ground state to be in accordance to Boltzmann's distribution:
\begin{equation}
    \frac{N_{\ket{\pm3/2}_{\mathrm{g}}}}{N_{\ket{\pm1/2}_{\mathrm{g}}}}=\mathrm{exp}\left(\frac{|2D|}{{k_{\rm{B}}}T_{\mathrm{Ruby}}}\right),    
\end{equation}
where $N_{{\ket{\pm1/2}_{\mathrm{g}}}}$ and $N_{{\ket{\pm3/2}_{\mathrm{g}}}}$ are the spin populations in states ${\ket{\pm1/2}_{\mathrm{g}}}$ and ${\ket{\pm3/2}_{\mathrm{g}}}$, and $T_{\mathrm{Ruby}}$ is the local lattice temperature.

\begin{figure}[!t]
	\centering 
		\includegraphics[width=\columnwidth]{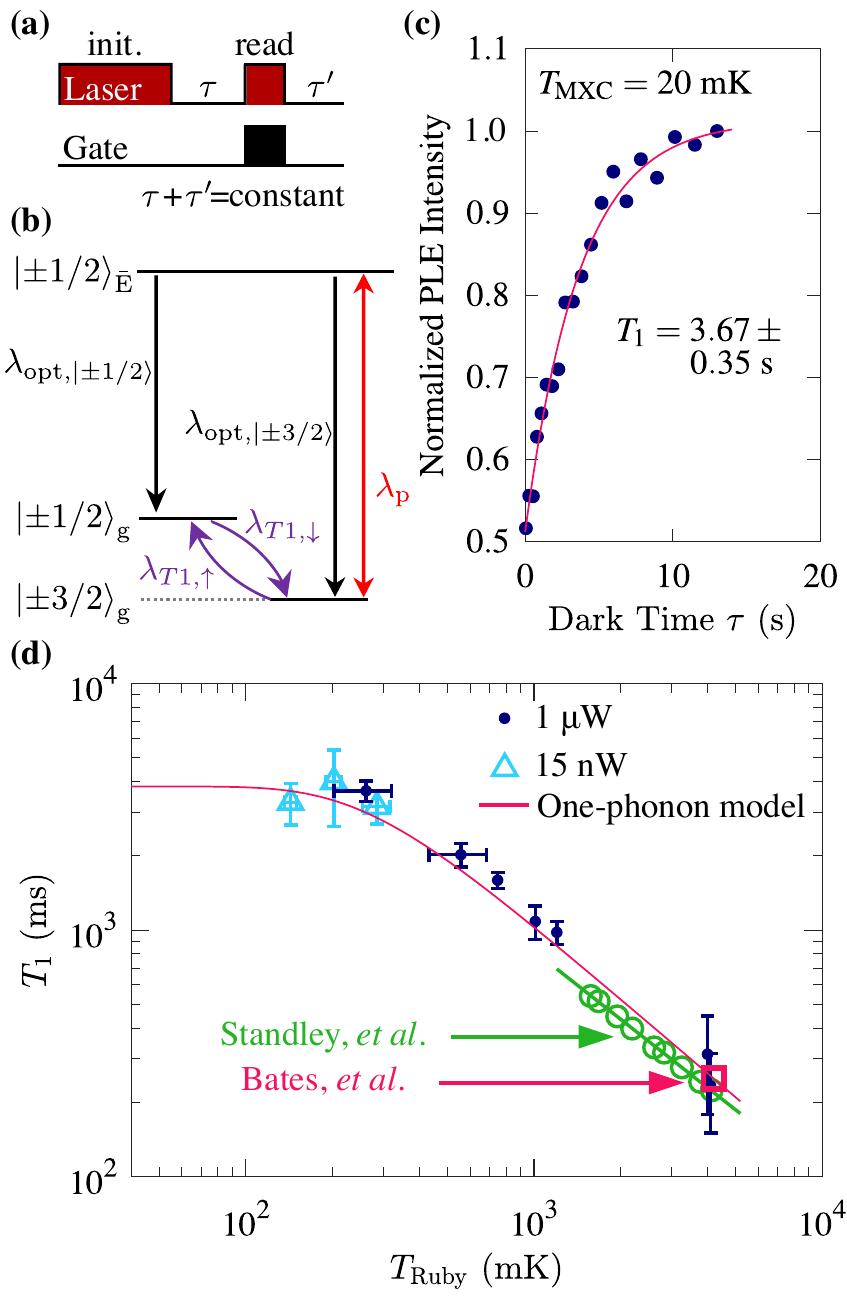}
		\caption{
		(a) Pulse sequence for the all-optical $T_{1}$ measurement. 
		(b) Rate model and transition rates describing the electron spin dynamics of the $\mathrm{Cr}^{3+}$ ground and excited state. See Appendix~\hyperref[App_B]{B} for more detail. 
		(c) $T_{1}$ spin decay curve measured at $T_{\rm{MXC}} = 20$~mK (base), $B_{0}=0$~T, and 1~$\upmu$W laser excitation power. 
		(d) Temperature dependence of $T_{1}$ as a function of lattice temperature $T_{\mathrm{Ruby}}$. The solid circles correspond to data measured with 1~$\upmu$W laser power, and the open triangles are confirmation measurements with 15~nW laser power for low temperatures. The open circles and the open square are literature values from Standley, \textit{et al.}~\cite{Standley1965} and Bates, \textit{et al.}~\cite{Bates1969}, respectively, shown for comparison. The literature value from Bates, \textit{et al.}~\cite{Bates1969} is additionally used to infer the $T_{1}$ at lower temperature assuming the one-phonon process described by Eq.~\ref{eqn:t1_theory} (solid line). 
		}
		\label{figure_spin_t1}
\end{figure}

The measurement of $T_{1}$ requires a deviation from the thermal equilibrium (initialization). This is achieved by optically pumping the $\mathrm{\ket{\pm3/2}_{g}}\leftrightarrow\bar{\mathrm{E}}$ transition to deplete $\mathrm{\ket{\pm3/2}_{g}}$ of electrons and cycle them into $\mathrm{\ket{\pm1/2}_{g}}$. We optically pump the system until a steady-state population has been reached. At 1~$\upmu$W of laser power, we measure the initialization time constant to be $0.65\pm0.13$~s [see Fig.~\ref{figure_S1}(a,b) in Appendix~\hyperref[App_A]{A}]. Hence, by $\sim3$~s we achieve steady-state, and based on the PL intensities, we transfer $\sim45$\% of the thermal $\mathrm{\ket{\pm3/2}_{g}}$ population to $\mathrm{\ket{\pm1/2}_{g}}$. 

The initialization pulse is the first pulse in the $T_{1}$ measurement sequence shown in Fig.~\ref{figure_spin_t1}(a). This is followed by a dark time $\tau$ of variable length in which the spins relax. A second laser pulse in resonance with the $\mathrm{\ket{\pm3/2}_{g}}\leftrightarrow\bar{\mathrm{E}}$ transition is used to read out the ground state spin population $N_{\mathrm{\ket{\pm3/2}_{g}}}$. We determine that the readout pulse length needs to be $\leq 0.1$~s in order to ensure $N_{\mathrm{\ket{\pm3/2}_{g}}}$ to be proportional to the total PL counts, which is achieved by measuring the spin signal as a function of optical pumping time in Fig.~\ref{figure_S2}(b) in Appendix~\hyperref[App_C]{C}, and verifying this with rate simulations based on the model shown in Fig.~\ref{figure_spin_t1}(b). The optical decay rate from the $\bar{\rm{E}}$ excited state to the $^4\rm{A}_2$ ground state $\lambda_{\rm{opt}}=\lambda_{\rm{opt},\ket{\pm1/2}}+\lambda_{\rm{opt},\ket{\pm3/2}}$ is determined by measuring the optical decay time $T_{1,\rm{opt}}=3.6\pm0.2$~ms [see Fig.~\ref{figure_S1}(e)]. The spin relaxation rates $\lambda_{T_1,\downarrow}$ and $\lambda_{T_1,\uparrow}$ have the following two relationships:  
\begin{equation}
    \label{eqn:t1_rates}
    \begin{gathered}
    \lambda_{T_1,\downarrow}+\lambda_{T_1,\uparrow}=T^{-1}_{1},
    \end{gathered}
\end{equation}
\begin{equation}
    \label{eqn:t1_ratesBoltzmann}
    \begin{gathered}
    \frac{\lambda_{T_1,\downarrow}}{\lambda_{T_1,\uparrow}} = \mathrm{exp}\left(\frac{|2D|}{k_{\rm{B}}T_{\mathrm{Ruby}}}\right).
    \end{gathered}
\end{equation}
The pump rate $\lambda_{\mathrm{p}}$ into the excited state is laser power dependent when resonantly driving the $\mathrm{\pm\ket{3/2}_{g}}\leftrightarrow\bar{\mathrm{E}}$ transition, and can be determined based on optical pumping decay curves. 

It is sufficient to only collect the PL emitted over the readout period, hence we set the `gate' channel of our SPAD to only enable counting during the readout period [see Fig.~\ref{figure_spin_t1}(a)]. A final consideration for performing the $T_{1}$ measurement is to ensure that the duty cycle of the laser, and therefore the average laser power, remains equal across all measurements in order to keep the sample at constant temperature. This is achieved by adding an additional dark time $\tau'$, as shown in Fig.~\ref{figure_spin_t1}(a), such that $\tau+\tau'$ is constant. 
At the base temperature of the dilution refrigerator ($T_{\rm{MXC}}=20$~mK), we measure $T_1=3.67\pm0.35$~s as shown in Fig.~\ref{figure_spin_t1}(c). 

Fig.~\ref{figure_spin_t1}(d) shows the measured $T_{1}$ at varying lattice temperatures $T_{\mathrm{Ruby}}$. The  $T_{\mathrm{Ruby}}$ of the measurements at 15~nW laser power (light blue triangles) were determined in Sec.~\ref{sec_thermometry}. Further $T_{1}$ measurements were performed with 1~$\upmu$W laser power (dark blue circles), and the corresponding $T_{\mathrm{Ruby}}$ were determined using similar methods to Sec.~\ref{sec_thermometry}, where at $T_{\mathrm{MXC}}=20$~mK and $T_{\mathrm{MXC}}=500$~mK, we calculate $T_{\mathrm{Ruby}}=262\pm59$~mK and $T_{\mathrm{Ruby}}=560\pm130$~mK, respectively. For $T_{\mathrm{MXC}}>500$~mK, we assume that $T_{\mathrm{MXC}}\approx T_{\mathrm{Ruby}}$. Assuming that the $T_{1}$ relaxation is dominated by a one-phonon (direct) process, we expect the following well-known relationship:
\begin{equation}
    \label{eqn:t1_theory}
    \begin{gathered}
    T_{1}^{-1} = A~\mathrm{coth}\left(\frac{|2D|}{2k_{\rm B}T_{\mathrm{Ruby}}}\right),
    \end{gathered}
\end{equation}
where $A$ is the coefficient of spontaneous emission rate~\cite{Ho2018}. Previous experimental results have shown that $T_{1}$ is dominated by the one-phonon process for $\mathrm{Cr}^{3+}$ concentrations $<0.01\%$ at low temperatures~\cite{Donoho}. At higher concentrations, cross-relaxation and exchange interaction between pairs of $\mathrm{Cr}^{3+}$ ions may deviate the temperature dependence of $T_{1}$ from the one-phonon process. Fig.~\ref{figure_spin_t1}(d) also shows $T_{1}$ as a function of temperature in a study by Standley \textit{et al.}~\cite{Standley1965} (open circles) supporting the one-phonon process relationship $>1.6$ K, shown for comparison. The discrepancy with our measured values arises from the fact that Ref.~\cite{Standley1965}'s data were measured with $B_{0}=330$~mT parallel to the crystal axis, across the pure $\ket{\pm1/2}$ ground state transition, and at a concentration of 0.017\%, different from our experimental condition. To conform to our conditions, we consider the study by Bates \textit{et al.}~\cite{Bates1969}, which has shown that at $B_{0}=0$ T and 4.2 K, the spin relaxation is limited to $T_{1}=250\pm20$ ms for $\mathrm{Cr}^{3+}$ concentrations $<0.02\%$ (open square). This result provides us with the spontaneous emission rate $A=0.252$ $\mathrm{s}^{-1}$ for Eq.~\ref{eqn:t1_theory}, and as our sample concentration of $0.005\%$ falls well below the mentioned thresholds, the temperature dependence of $T_{1}$ can be completely predicted below 4.2 K by the one-phonon model (Eq.~\ref{eqn:t1_theory}), and is shown in Fig.~\ref{figure_spin_t1}(d) as a solid line. Our measured $T_{1}$ is in excellent correspondence.     

Early results have found that the spin relaxation at low temperatures can ultimately depend on a number of parameters, such as growth method, chromium concentration, measurement frequency, chemical purity, and magnetic field strength and angle~\cite{Standley1965,Mason1967}.

\begin{figure*}
	\centering 
		\includegraphics[width=\textwidth]{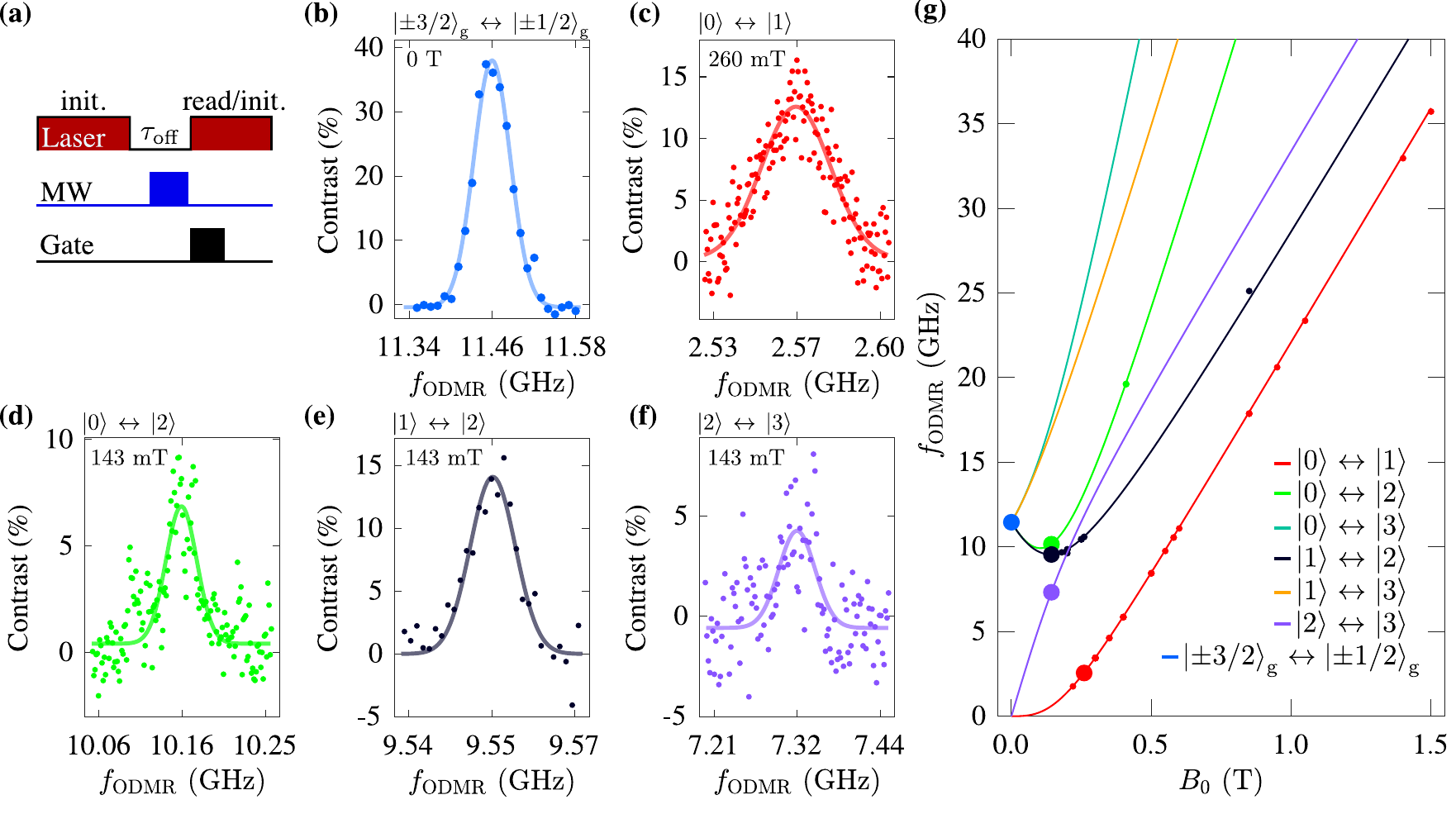}
		\caption{
		(a) Pulse sequence for implementing pulsed ODMR. The initialization laser pulse is 3~s long, and the second laser pulse is also 3~s long and acts as a readout and re-initialization pulse. Within the $\mathrm{\tau}_{\mathrm{off}}=10$~ms dark time, a MW pulse of 10~$\upmu$s - 1~ms length is applied, aligned to occur just before the readout pulse. The SPAD is gated to detect photons for the first 0.1~s of readout. The laser excitation power is 1~$\upmu$W - 10~$\upmu$W.
		(b-f) ODMR transitions with linewidths [full-width half-maximum (FWHM)] (b) $62.0\pm4.3$~MHz, (c) $38.2\pm4.7$~MHz, (d) $32.7\pm4.9$~MHz, (e) $8.8\pm2.0$~MHz, and (f) $52\pm17$~MHz. 
		(g) All ODMR transition frequencies measured, superimposed with expected transition frequencies given by Eq.~\ref{eqn:hamiltonian}. The data points obtained from (b-f) are emphasized with larger circles.   
		}
		\label{figure_odmr}
\end{figure*}

\section{Optically-detected magnetic resonance}\label{sec_odmr}
In this section we perform optically-detected electron spin resonance in between the spin states of the $^4\rm{A}_2$ ground state. The pulse sequence to implement ODMR is shown in Fig.~\ref{figure_odmr}(a). We set our Ti:Sa laser in resonance with a particular spin state $\ket{i}$ and the optically excited $\ket{\pm1/2}_{\bar{\mathrm{E}}}$ state. By applying a laser pulse at this wavelength, we deplete the electron population in $\ket{i}$ via optical cycling into other states, just as we did with the laser initialization pulse in Sec.~\ref{sec_t1}. In the dark time $\tau_{\rm{off}}$ after the laser pulse, a MW pulse is delivered to the sample via the PCB antenna mounted above the ruby sample. As discussed in Sec.~\ref{sec_setup}, the antenna creates an oscillating magnetic field $B_{1} \parallel B_{0}$. If the MW pulse is resonant with a transition between $\ket{i}$ and any other spin level $\ket{j}$, we satisfy the resonance condition and re-introduce some population into $\ket{i}$. Finally, we use a readout laser pulse, as in Sec.~\ref{sec_t1}, to measure the population in $\ket{i}$ at the end of the MW pulse. 

We perform this measurement across various transitions, for magnetic fields $0\leq B_{0}\leq1.5$~T and transition frequencies up to 36~GHz, with some examples shown in Fig.~\ref{figure_odmr}(b-f). The ODMR contrast measured is the percentage change in the $\ket{i}$ population due to a resonant MW pulse, compared to the initialized population. The positive contrast values in all scans indicate that our initialization pulse is depleting the state $\ket{i}$ to a population that is always less than the population of $\ket{j}$. 

In Appendix~\hyperref[App_D]{D} we provide time evolution simulations of the magnetically-driven transitions $\ket{0}\leftrightarrow\ket{1}$, $\ket{0}\leftrightarrow\ket{2}$, and $\ket{1}\leftrightarrow\ket{2}$ under the condition $B_{0} \perp$ c-axis, as in our experiments. The simulations show that with the $B_{1}$ direction perfectly aligned with either the c-axis or $B_{0}$, some transitions are theoretically forbidden, and not all transitions in Fig.~\ref{figure_odmr}(b-f) should be observable with our experimental setup (see also Ref.~\cite{Bois1959}). We conclude therefore, that our $B_{1}$ field must also have a small component along the c-axis. 

In Fig.~\ref{figure_odmr}(g) we mark all observed transitions frequencies onto the theoretical transition frequencies from Eq.~\ref{eqn:hamiltonian}, in good correspondence [data points from Fig.~\ref{figure_odmr}(b-f) are marked as larger circles]. Noticeable here is that the ground state Hamiltonian hosts two clock transitions: $\ket{0}\leftrightarrow\ket{2}$ at $B_{0}=103$~mT and $\ket{1}\leftrightarrow\ket{2}$ at $B_{0}=143$~mT. The latter one was observed [see Fig.~\ref{figure_odmr}(e)] and exhibits the minimum linewidth of $8.8\pm2.0$~MHz of all transitions measured. Finally, it is worth noting that the broadening of the zero-field ODMR shown in Fig.~\ref{figure_odmr}(b) is predominantly due to the dispersion of the zero-field splitting parameter $D$ in Eq.~\ref{eqn:hamiltonian}, due to variations in local strain near $\mathrm{Cr}^{3+}$ ions~\cite{Kirkby_1968}.

\section{Conclusion}
In this paper we perform spin characterization of $\rm{Cr}^{3+}$ ions in $\rm{Al}_2{O}_3$ (ruby) at ultra-low temperatures $T \ll 1$~K. We optically read out the ${\rm S}=3/2$ spin state populations using photoluminescence excitation spectroscopy with phonon sideband detection. We conduct thermometry by measuring the relative populations as a function of magnetic field, and demonstrate that a lattice temperature as low as $T_{\rm{Ruby}}=143\pm7$~mK can be achieved under continuous laser excitation. This allows us to perform spin relaxation measurements in the temperature range $T \ll 1$~K. We measure a maximum spin relaxation time of $T_1=3.67\pm0.35$~s and confirm that the spin relaxation dynamics are governed by a direct, one-phonon process for ultra-low temperatures. Furthermore, we perform optically-detected magnetic resonance in between the $^4\rm{A}_2$ ground state spin levels, observing transitions up to 36~GHz.

\section*{ACKNOWLEDGMENTS}
We acknowledge support from the Australian Research Council (LE160100069). A.L. acknowledges support through the UNSW Scientia Program. T.S. acknowledges support from the TUM Graduate School.

\section*{Appendix A: Determining Pump Initialization Time and Optical \texorpdfstring{$\mathbf{\textit{T}_{1}}$}{T1}}\label{App_A}

\begin{figure*}
	\centering 
		\includegraphics[width=\textwidth]{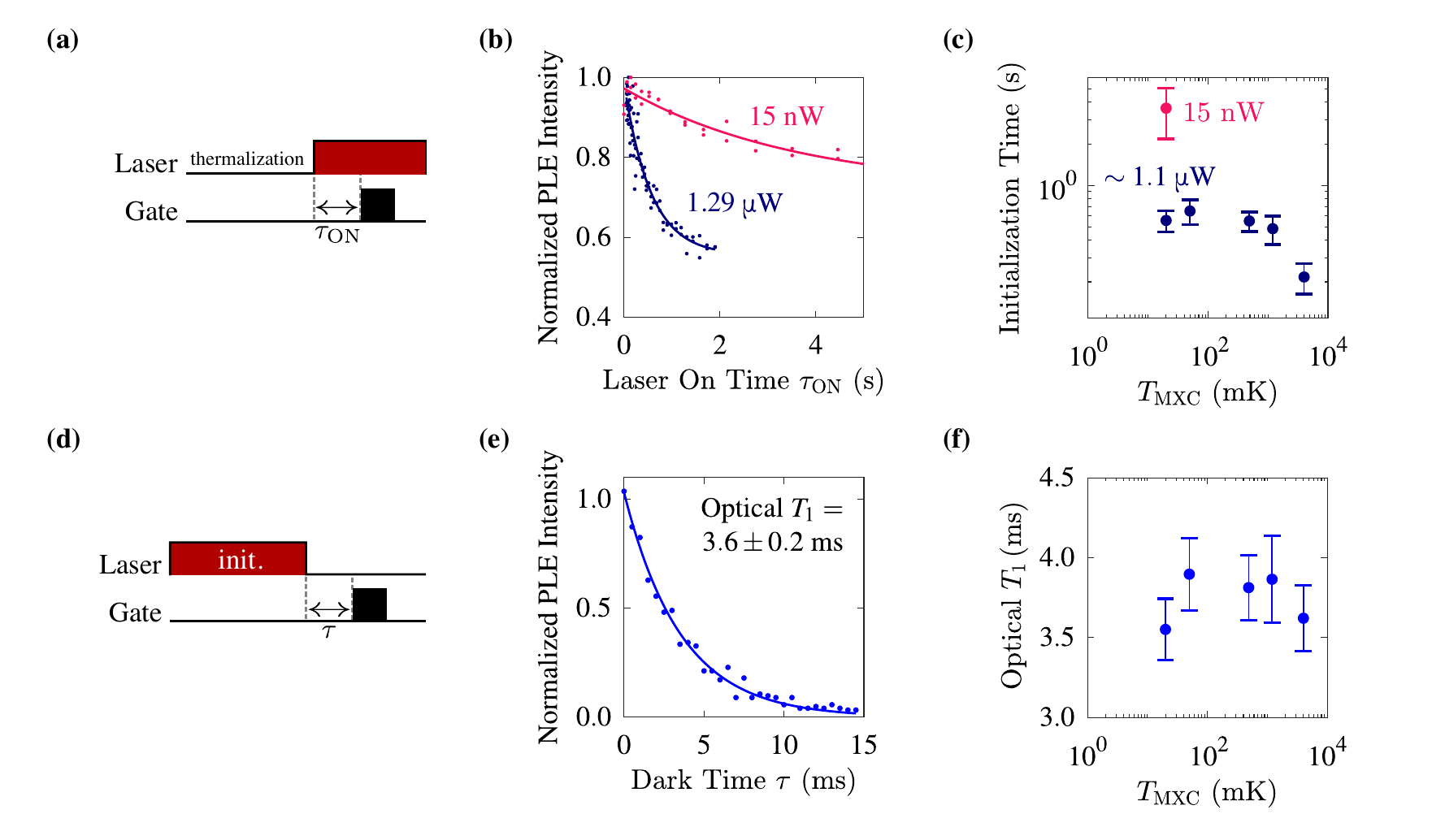}
		\caption{
		(a) Pulse sequence for measuring the initialization time constant.
		(b) PL signal as a function of laser on time $\tau_{\rm{ON}}$ before detecting photons at $T_{\mathrm{MXC}}=20$ mK.
		(c) Extracted initialization time constants as a function of temperature $T_{\rm{MXC}}$.
		(d) Pulse sequence for measuring the optical relaxation time $T_{1,\rm{opt}}$.
		(e) Optical relaxation time $T_{1,\rm{opt}}$ measurement at $T_{\mathrm{MXC}}=20$ mK. 
		(f) $T_{1,\rm{opt}}$ dependence on temperature $T_{\rm{MXC}}$.
		}
		\label{figure_S1}
\end{figure*}

The measurement of the spin relaxation time $T_{1}$ (see Sec.~\ref{sec_t1}) and ODMR (see Sec.~\ref{sec_odmr}) requires an initialization laser pulse to polarize the ground state spins away from their thermal (Boltzmann) equilibrium. This initialization pulse must be long enough to reach a steady state in population, allowing us to start with the same initial ground state population for every repetition of the measurement sequences shown in Fig.~\ref{figure_spin_t1}(a) and Fig.~\ref{figure_odmr}(a). Fig.~\ref{figure_S1}(a) depicts the pulse sequence that we use to determine this initialization time. After a long (20~s $\gg T_{1}$) dark time to allow the spin populations to return to their thermal equilibrium, we turn on the laser and monitor the PL as a function of on-time $\tau_{\rm{ON}}$ by shifting the gating pulse for the SPAD to later times. In Fig.~\ref{figure_S1}(b) we plot the reduction in PL intensity as a function of laser on-time. The graphs demonstrate how quickly and efficiently the laser depopulates the state in resonance for 15~nW (red points) and 1.3~$\upmu$W (blue points) laser power. The solid lines are exponential fits to the data indicating an initialization time constant of $3.6\pm1.4$~s for 15~nW and $0.55\pm0.10$~s for 1.3~$\upmu$W laser power, based on which we set the initialization pulse time to be 10~s and 3~s, respectively. We also plot the temperature dependence of the initialization time constant in Fig.~\ref{figure_S1}(c). At low temperatures it stays fairly constant and then starts reducing for $T_{\rm{MXC}}>0.5$~K. This is due to the fact that the initialization time is determined by both the pump rate $\lambda_{\mathrm{p}}$ as defined in Fig.~\ref{figure_spin_t1}(b), as well as the spin relaxation rate $T_{1}^{-1}$. For $\lambda_{\mathrm{p}}$ approaching 0 Hz, the initialization rate will approach $T_{1}^{-1}$. For $\lambda_{\mathrm{p}}$ being comparable to $T_{1}^{-1}$, which we show to be true in Appendix~\hyperref[App_B]{B}, a change in $T_{1}$ will have an observable effect on the overall initialization time.

The pulse sequence for measuring the optical relaxation time $T_{1,\rm{opt}}$ is depicted in Fig.~\ref{figure_S1}(d). $T_{1,\rm{opt}}$ is found by optically pumping the system with a laser pulse, then turning it off, and monitoring the PL as a function of wait time $\tau$. This is again realized by shifting the gating pulse for the SPAD to later times. We plot the PL decay curve in Fig.~\ref{figure_S1}(e), where the solid line is an exponential fit to the data. We find $T_{1,\rm{opt}}=3.6\pm0.2$~ms in this measurement, with no observable dependence on temperature [see Fig.~\ref{figure_S1}(f)]. 
The optical $T_{1,\rm{opt}}$ in ruby is well-known in literature, going back to the first measurements by Becquerel with his phosphoroscope in 1867~\cite{Becquerel1867}. It stays relatively constant for temperatures below 300~K~\cite{Zhang1993,Chandler2006}, and is in good agreement with our data.

\section*{Appendix B: Rate Model}\label{App_B}
We refer to the rate model and transition rates depicted in Fig.~\ref{figure_spin_t1}(b) of Sec.~\ref{sec_t1}, which are applicable for $B_0=0$~T only. For all experiments performed in Sec.~\ref{sec_t1} and Sec.~\ref{sec_odmr} at $B_0=0$~T, only the transition $\ket{\pm3/2}_{\mathrm{g}}\leftrightarrow\ket{\pm{1/2}}_{\bar{\mathrm{E}}}$ has been resonantly excited with a laser. Hence, the rate model only includes a single pump rate $\lambda_{\mathrm{p}}$ (units of Hz) across this transition. The time evolution of the model is simulated in Matlab using Runge-Kutte methods, using the following rate equations:
\begin{widetext}
    \begin{equation}
        \label{eqn:rates}
        \begin{pmatrix}
        {\dot{N}_{\ket{\pm{1/2}}_{\bar{\mathrm{E}}}}} \\
        {\dot{N}_{\ket{\pm{1/2}}_{\mathrm{g}}}} \\
        {\dot{N}_{\ket{\pm{3/2}}_{\mathrm{g}}}} 
        \end{pmatrix}
        =
        \begin{pmatrix}
        -\left(\lambda_{\mathrm{p}}+\lambda_{\mathrm{opt,\ket{\pm1/2}}}+\lambda_{\mathrm{opt,\ket{\pm3/2}}}\right) && 0 && \lambda_{\mathrm{p}} \\
        \lambda_{\mathrm{opt,\ket{\pm1/2}}} && -\lambda_{T1,\downarrow} && \lambda_{T1,\uparrow} \\
        \lambda_{\mathrm{p}}+\lambda_{\mathrm{opt,\ket{\pm3/2}}} && \lambda_{T1,\downarrow} && -\left(\lambda_{\mathrm{p}}+\lambda_{T1,\uparrow}\right)
        \end{pmatrix}
        \begin{pmatrix}
        N_{\ket{\pm{1/2}}_{\bar{\mathrm{E}}}}\\
        N_{\ket{\pm{1/2}}_{\mathrm{g}}} \\
        N_{\ket{\pm{3/2}}_{\mathrm{g}}} 
        \end{pmatrix}.
    \end{equation}
\end{widetext}

\begin{figure*}
	\centering 
		\includegraphics[width=\textwidth]{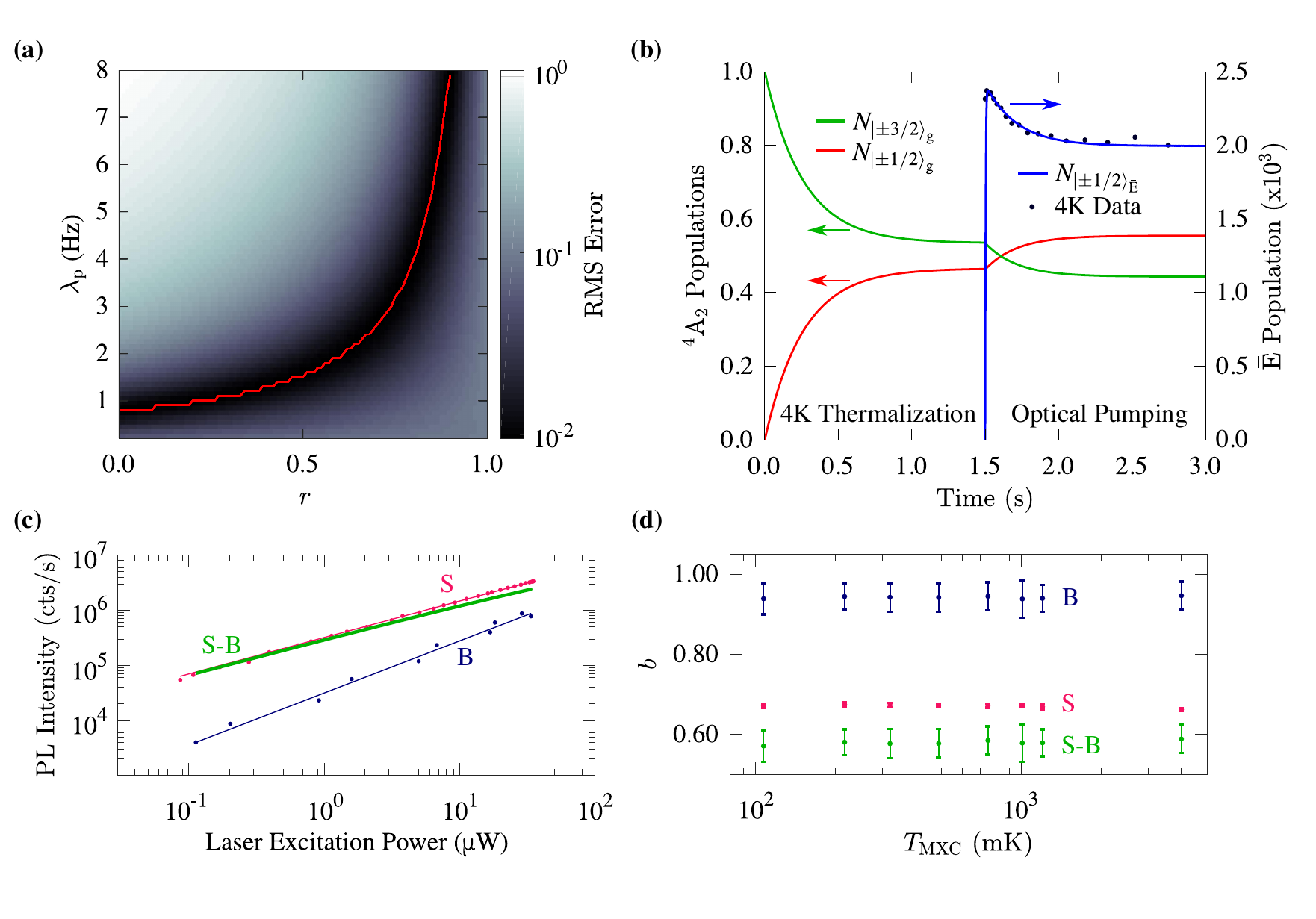}
		\caption{
		(a) Root-mean-square errors of best-fits to the 1~$\upmu$W, 4~K spin initialization curve as a function of laser pump rate $\lambda_{\rm{p}}$ and branching ratio $r$. The red line indicates the curve of the lowest fitting error.
		(b) Simulation of the spin initialization measurement using the rate model Eq.~\ref{eqn:rates}, with $\lambda_{\mathrm{p}}=1.5$~Hz and $r=0.5$. The populations $N_{\ket{\pm{1/2}}_{{\mathrm{g}}}}$ (red line), $N_{\ket{\pm{3/2}}_{{\mathrm{g}}}}$ (green line), and $N_{\ket{\pm{1/2}}_{\bar{\mathrm{E}}}}$ (blue line), are superimposed with the experimental data in arbitrary units (black circles). 
		(c) Excitation power dependence at $T_{\mathrm{MXC}}=$~20mK of the PLE signal ``S'' (red circles), and the background ``B'' (blue circles) with fits in the form ${a}{x}^{b}$, where $x$ is the laser excitation power. The green line shows the expected signal with the background subtracted ``S-B''.
		(d) Extracted exponential $b$ from the excitation power dependence in (c) performed at varying $T_{\rm MXC}$.
		}
		\label{figure_S2}
\end{figure*}

$N_{\ket{i}}$ is the electron population of the level $\ket{i}$, ${\dot{N}_{\ket{i}}}=\frac{\mathrm{d}N_{\ket{i}}}{\mathrm{d}t}$, and $\left(N_{\ket{\pm{1/2}}_{\bar{\mathrm{E}}}}+N_{\ket{\pm{1/2}}_{\mathrm{g}}}+N_{\ket{\pm{3/2}}_{\mathrm{g}}}\right)=1$. The individual spontaneous emission rates $\lambda_{\mathrm{opt},\ket{\pm1/2}}$ and $\lambda_{\mathrm{opt},\ket{\pm3/2}}$ are not determined in this study, however, we set $T_{1,\mathrm{opt}}=\lambda_{\mathrm{opt},\ket{\pm1/2}}+\lambda_{\mathrm{opt},\ket{\pm3/2}}=\left(3.6~{\rm ms}\right)^{-1}$ ms. The ground state spin $T_{1}$ parameters can be calculated based on Eq.~\ref{eqn:t1_rates} and~\ref{eqn:t1_ratesBoltzmann}. Without the knowledge of the $T_{1}$ parameters determined in this paper, we can perform a simulation at 4~K based on the literature value of $T_{1}=0.26$~ms at $B_0=0$~T, as in Ref.~\cite{Bates1969}. Based only on the simple model in Fig.~\ref{figure_spin_t1}(b), the unknown variables are then the pump rate $\lambda_{\mathrm{p}}$ and the branching ratio 
\begin{equation}
    \label{eqn:decayratio}
   r=\frac{\lambda_{\mathrm{opt},\ket{\pm3/2}}}{(\lambda_{\mathrm{opt},\ket{\pm3/2}}+\lambda_{\mathrm{opt},\ket{\pm1/2}})},
\end{equation}
which is the probability of spontaneous emission into the $\ket{\pm3/2}_{\mathrm{g}}$ state. 

We first simulate the spin initialization time measurement [see the pulse sequence in Fig.~\ref{figure_S1}(a)]. We sweep the parameters $r$ and $\lambda_{\mathrm{p}}$, and for each pair of values we perform a best-fit to the experimental spin initialization decay curve at 4~K and 1~$\upmu$W of laser excitation power. Fig.~\ref{figure_S2}(a) shows the root-mean-square error of the performed fits. A range of values can adequately fit the 4~K initialization decay curve (red line), illustrating that there is no unique pair of values. Early results have determined that $r\approx0.55$~\cite{Sugano1958b,Nelson1965}, hence based on the solutions shown in Fig.~\ref{figure_S2}(a), we expect the pump rate $\lambda_{\mathrm{p}}\approx1.7$ Hz at $T_{\mathrm{MXC}}=4$ K and 1~$\upmu$W laser power.

Fig.~\ref{figure_S2}(b) depicts the output of the simulation for the case of $\lambda_{\mathrm{p}}=1.5$~Hz and $r=0.5$ (any pair of values on the red curve in Fig.~\ref{figure_S2}(a) would result in a similar simulation output). Assuming we begin with zero electron occupancy in the $\ket{\pm1/2}_{\mathrm{g}}$ state (red line) and complete electron occupancy in the $\ket{\pm3/2}_{\mathrm{g}}$ state (green line) with the laser off ($\lambda_{\mathrm{p}}=0$~Hz), the ground state populations converge to thermal equilibrium in accordance with their Boltzmann factors at 4~K at the rate of $1/T_{1}$. At 1.5~s into the simulation, we set $\lambda_{\mathrm{p}}>0$~Hz and track the population in all three states. At the end of 3~s, the ground state populations converge to a new steady state distribution. At all times, $N_{\ket{\pm{1/2}}_{\bar{\mathrm{E}}}}$ is assumed to be proportional to the optical emission. For the values chosen, the simulation is in good agreement with the measured data at 4~K (black circles).

\section*{Appendix C: Effect of Spin Pumping on Measurements}\label{App_C}

In addition to the PLE signal being proportional to $N_{\ket{\pm{1/2}}_{\bar{\mathrm{E}}}}$, it is also proportional to $N_{\ket{\pm{3/2}}_{{\mathrm{g}}}}$ at any point in time the laser is pumping the $\ket{\pm3/2}_{\mathrm{g}}\leftrightarrow\ket{\pm{1/2}}_{\bar{\mathrm{E}}}$ transition. For the solutions obtained in Fig.~\ref{figure_S2}(b), we confirm that regardless of the gate pulse duration chosen, the integrated PL will have some linear relationship to the population in $\ket{\pm3/2}_{\mathrm{g}}$ prior to the gate pulse (Pearson-$r>0.99$). This is sufficient to observe the spin $T_{1}$ decay in Fig.~\ref{figure_spin_t1}. However, in the case of short gate pulses, the PL observed is more directly proportional (i.e. linear with no offset, where the offset represents PL that is independent of the $\ket{\pm3/2}_{\mathrm{g}}$ population prior to the gate pulse), while longer gate pulses contain increasingly large offsets. This can be clearly seen in Fig.~\ref{figure_S2}(b), where the initial (thermal) spin population is quickly dissipated by the optical cycling. We choose to reduce the contribution of the spin-signal-independent contribution and maximize the proportion of the spin-signal-dependent contribution, by only collecting PL from the beginning of the readout pulse. We, therefore, choose a gate pulse time of $0.1$~s for measurements performed at 1~$\upmu$W, and following a similar analysis, $0.5$~s for measurements at 15~nW.

In Fig.~\ref{figure_S2}(c), we plot the power dependence of the PLE intensity as a function of laser excitation power on a double-logarithmic scale for $T_{\rm MXC}=4$~K and $B_0=0$~T. The green curve (``S'') shows the total collected signal when the laser is in resonance with the $\ket{\pm3/2}_{\mathrm{g}}\leftrightarrow\ket{\pm{1/2}}_{\bar{\mathrm{E}}}$ transition. The blue curve (``B'') corresponds to the background signal collected when the laser is not in resonance with any of the discrete ${\rm Cr}^{3+}$ transitions. 
We fit both S and B to power law dependencies and extract their exponents. For B, we obtain an exponent of $0.95\pm0.04$, indicating a linear power dependence. For S, the signal saturates at high excitation powers, resulting in a sublinear dependence with an exponent of $0.59\pm0.04$. 
Based on the fits, we also plot the background-adjusted signal ``S-B'' (green curve).

In Fig.~\ref{figure_S2}(d) we plot the extracted exponents as a function of temperature $T_{\rm MXC}$. Notably, the exponent of the background is temperature-independent and linear in power, giving us validation for removing the background from all data before analysis.

\newpage

\section*{Appendix D: Simulations of Rabi Oscillations}\label{App_D}

\begin{figure*}
	\centering 
		\includegraphics[width=\textwidth]{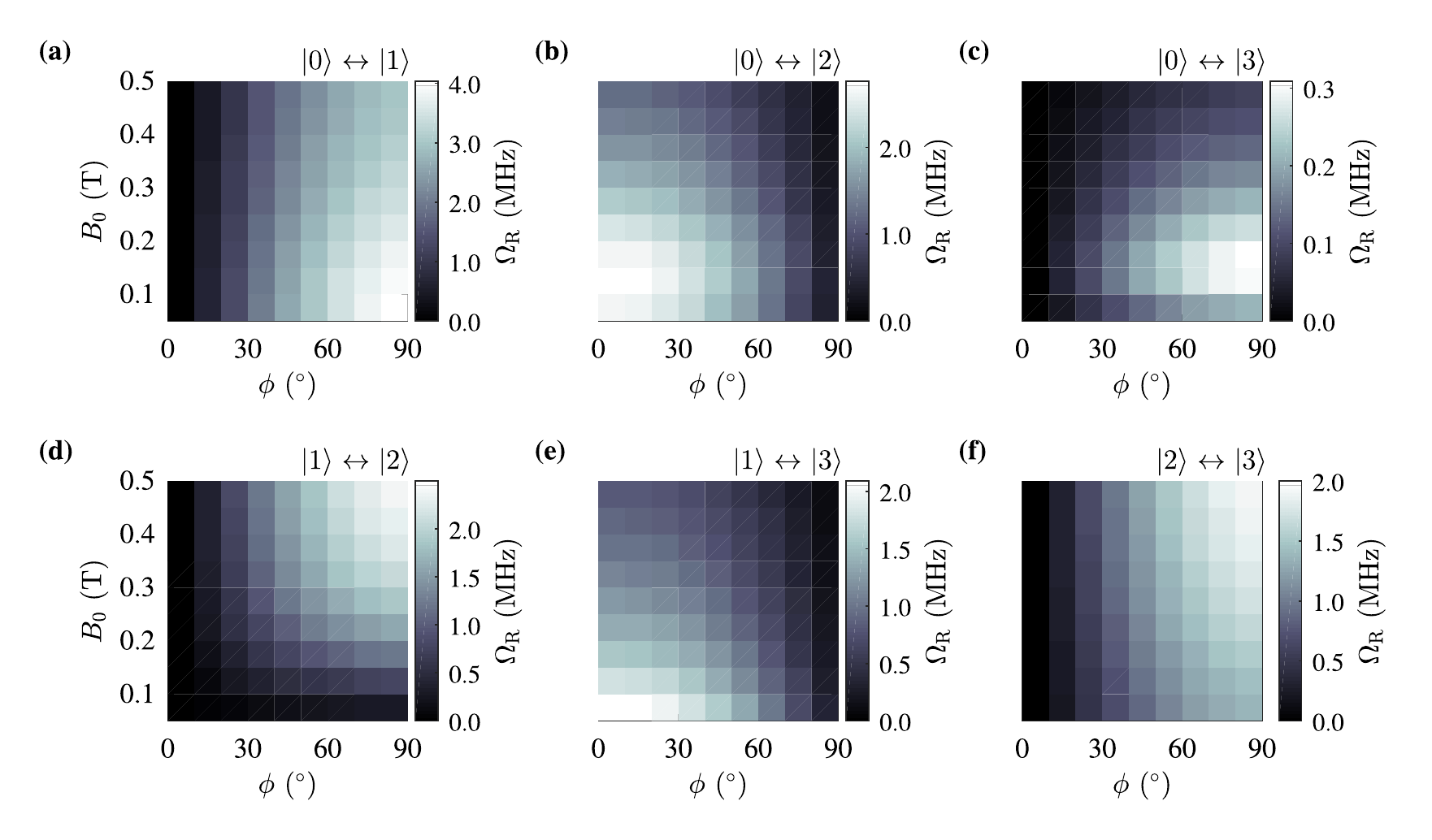}
		\caption{
		(a-f) Rabi frequencies $\Omega_{\rm{R}}$ extracted from time evolution simulations for transitions (a) $\ket{0}\leftrightarrow\ket{1}$, (b) $\ket{0}\leftrightarrow\ket{2}$, (c) $\ket{0}\leftrightarrow\ket{3}$, (d) $\ket{1}\leftrightarrow\ket{2}$, (e) $\ket{1}\leftrightarrow\ket{3}$, and (f) $\ket{2}\leftrightarrow\ket{3}$ as a function of $B_0$ strength and $B_1$ direction as specified by $\phi$.
		}
		\label{figure_S4}
\end{figure*}

In order to examine the efficiency at which we can perform magnetic resonance on various transitions, we perform numerical simulations in Matlab. We take the ground state Hamiltonian ${\cal H}_0$ (Eq.~\ref{eqn:hamiltonian}), and apply $B_{0}\perp$ c-axis, along the x-axis without loss of generality. We add a time-dependent part ${\cal H}_1$ that describes the effect of a linearly polarized microwave field with strength $B_{1}=100$~$\upmu$T:
\begin{equation}
    \label{eqn:H1}
    \begin{gathered}
    {\cal H}_1=\left(g_{\parallel}~\mathrm{sin}(\phi)~\mathrm{S}_{z}+g_{\perp}~\mathrm{cos}(\phi)~\mathrm{S_{x}}\right)~{\mu_{\rm B}}{B}_{1}\mathrm{sin}(\omega t),
    \end{gathered}
\end{equation}
where $\omega$ is the angular frequency of the driving field and $\phi$ is the angle between $B_{1}$ and $B_{0}$. Here $\phi=0^{\circ}$ corresponds to $B_{1} \parallel B_{0}$ along the x-axis, and $\phi=90^{\circ}$ corresponds to $B_{1} \parallel$ c-axis along the z-axis. The case of $B_1$ along the y-axis, i.e. $B_{1} \perp$ c-axis while also $B_{1} \perp B_{0}$, is not considered.

In Fig.~\ref{figure_S4} we plot the extracted Rabi frequencies as a function of $B_0$ magnitude (for $0.05~\mathrm{T}\leq B_{0} \leq 0.5$~T), and $B_1$ direction as specified by $\phi$ (for $0^{\circ} \leq \phi \leq 90^{\circ}$). The result for the transition $\ket{0}\leftrightarrow\ket{1}$ is shown in Fig.~\ref{figure_S4}(a), for $\ket{0}\leftrightarrow\ket{2}$ in Fig.~\ref{figure_S4}(b), for $\ket{0}\leftrightarrow\ket{3}$ in Fig.~\ref{figure_S4}(c), for $\ket{1}\leftrightarrow\ket{2}$ in Fig.~\ref{figure_S4}(d), for $\ket{1}\leftrightarrow\ket{3}$ in Fig.~\ref{figure_S4}(e), and for $\ket{2}\leftrightarrow\ket{3}$ in Fig.~\ref{figure_S4}(f). At $B_{0}=0$ T (not shown), irrespective of $\phi$, only the $\ket{0}\leftrightarrow\ket{2}$ and $\ket{1}\leftrightarrow\ket{3}$ transitions are allowed as there would be no off-diagonal entries in Eq.~\ref{eqn:hamiltonian} and as these transitions result in a spin change of 1. Based on the simulation results shown, for $B_{0}>0$ T and $\phi=0^{\circ}$, the $\ket{0}\leftrightarrow\ket{2}$ and $\ket{1}\leftrightarrow\ket{3}$ transitions are allowed, albeit with decreasing Rabi frequency as the increasing $B_{0}$ tilts the quantization axis of ${\cal H}_0$ away from $\perp B_{1}$ and towards $\parallel B_{1}$. Under the same conditions, all other transitions are forbidden.
In contrast, the case of $\phi=90^{\circ}$ yields a forbidden transition for the $\ket{0}\leftrightarrow\ket{2}$ and $\ket{1}\leftrightarrow\ket{3}$ case, and allowed transitions for all other cases. See also Ref.~\cite{Bois1959} for more details.

In main text Fig.~\ref{figure_odmr}(b-d), the transitions $\ket{0}\leftrightarrow\ket{1}$, $\ket{0}\leftrightarrow\ket{2}$, $\ket{1}\leftrightarrow\ket{2}$, $\ket{2}\leftrightarrow\ket{3}$ are observed, which is possible for intermediate values of $\phi$. In other words, in order to observe all ODMR transitions with $B_{0} \perp$ c-axis, it suffices to have the $B_{1}$ vector parallel to the plane defined by the $B_{0}$ vector and c-axis, but not aligned exactly to either.

\bibliography{mybib}

\end{document}